\documentclass[reprint,amsmath,amssymb,aps]{revtex4-2}
\usepackage{gensymb}
\usepackage[bottom]{footmisc}
\usepackage[utf8]{inputenc}
\usepackage{float}
\usepackage[english]{babel}
\addto\captionsenglish{}
\usepackage{color,epsfig}
\usepackage{array}
\usepackage{amsmath,amssymb}
\usepackage{natbib}
\usepackage{bbold}
\usepackage{dsfont}
\usepackage{graphicx}
\usepackage[normalem]{ulem} 
\usepackage[numbered]{matlab-prettifier}
\usepackage[autostyle]{csquotes}
\usepackage{comment}
\usepackage{bm}
\usepackage[pdftex,pdfborder={0 0 0}]{hyperref}

\begin{document}


\title{Sub-µm axial precision depth imaging with entangled two-colour Hong-Ou-Mandel microscopy}

\author{Cyril Torre$^{1,2}$, Alex McMillan$^{1}$, Jorge Monroy-Ruz$^{1}$ and Jonathan C.F. Matthews$^{1}$}
\affiliation{1.Quantum Engineering Technology Labs, H.H. Wills Physics Laboratory and Department of Electrical $\&$ Electronic Engineering, University of Bristol, Tyndall Avenue, BS8 1FD, United Kingdom.}%
\affiliation{2.Quantum Engineering Centre for Doctoral Training, H.H. Wills Physics Laboratory and Department of Electrical $\&$ Electronic Engineering, University of Bristol, Tyndall Avenue, BS8 1FD, United Kingdom}


\date{\today}


\begin{abstract}
The quantum interference of two wavelength-entangled photons overlapping at a beamsplitter results in an oscillating interference pattern. The frequency of the beat note is dependent on the wavelength separation of the entangled photons but is robust to wavelength scale perturbations that can limit the practicality of standard interferometry. 
Here we use two-colour entanglement interferometry to \textcolor{black}{evaluate the variation in thickness of a semi-transparent sample in combination with two-dimensional raster scanning}. 
The axial precision and the dynamic range of the microscope are actively controlled by adjusting the wavelength separation of the entangled photon pairs. Sub\nobreakdash-$\mu m$ precision is reported using up to \textcolor{black}{$12.3~nm$} of detuning and \textcolor{black}{$\sim10^4$ detected} photon pairs. 
\end{abstract}

\maketitle

\begin{figure}[!htp]
    \centering
    \includegraphics[scale = 0.25]{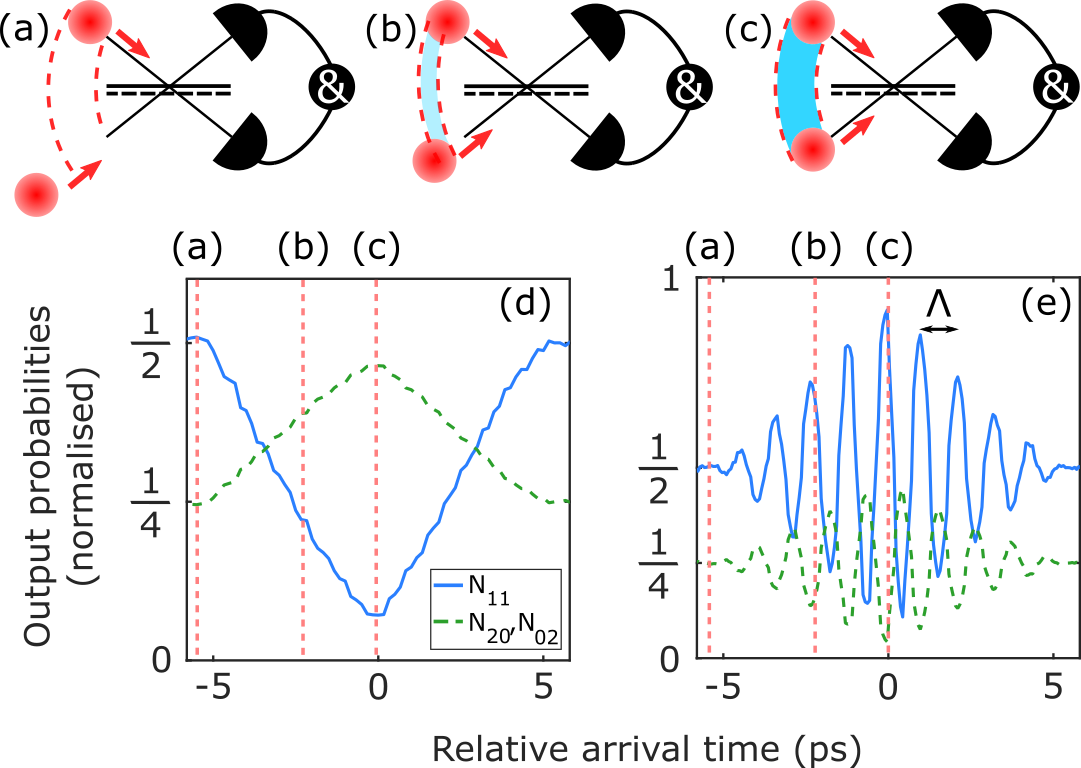}
    \caption{\textbf{HOM interference for degenerate and two-colour entangled photons.} (a--c) Photon pairs incident on a beamsplitter with zero (a), partial (b) and complete (c) temporal overlap. Typical resulting HOM interference patterns for degenerate (d) and two-colour entangled photons \textcolor{black}{with a frequency beat note of 0.90 THz} (e) are plotted, showing bunching ($N_{20}$, $N_{02}$ --- green dashed) and anti-bunching ($N_{11}$ --- blue) rates. Overlap conditions corresponding to cases (a--c) are highlighted. 
   The wavelength of the oscillating interference fringe $\Lambda$ (marked on (e)) is directly controlled by the frequency detuning ($\Delta\nu=1/\Lambda$) of signal and idler wavelengths.}
    \label{HOMCartoon}
\end{figure}

The Hong-Ou-Mandel (HOM) interference effect is routinely used in quantum photonics, for example to generate NOON states for quantum enhanced interferometry~\cite{Slussarenko2017Nov} and imaging~\cite{Ono2013Sep, Israel2014Mar}, for photonic two-qubit quantum logic gates~\cite{RalphCNOT, HofmannTakeuchiCZ, OBrienCNOTExpt} and it is the underpinning physical \textcolor{black}{phenomenon} of optical quantum computational advantage experiments~\cite{doi:10.1126/science.abe8770}. It is typically observed when two photons in separate temporal modes, but otherwise indistinguishable,
are made to overlap at a beamsplitter (Fig.~\ref{HOMCartoon}). As the relative arrival time $t$ of the two photons is reduced to zero, the detection rate of the photons exiting the beamsplitter \textcolor{black}{on separate paths} (anti-bunching) ideally drops to zero, while the \textcolor{black}{proportion of detection events with photons output on the same path} (bunching) increases. This can be used to measure a relative optical path length that induces $t$, which was demonstrated to subpicosecond precision in the seminal experiment by Hong, Ou and Mandel~\cite{HOMdip}. 

\textcolor{black}{For application of the HOM effect to} imaging and sensing, more recent demonstrations have extracted higher sensitivities to measure $t$ with attosecond precision~\cite{Ashley}, \textcolor{black}{and have utilised the effect} for polarisation measurements~\cite{HOMpola}, spectroscopy ~\cite{HOMspectroscopy}, multi-mode depth imaging~\cite{HOMmicroscopy} and refractive index measurement~\cite{HOMNice}. In these works, the Cram\'{e}r-Rao bound (CRB) \textcolor{black}{relates the variance~($\sigma^2_ t$) to the Fisher information~($\mathcal{F}(t)$) through the inequality} $\sigma^2_ t\geq 1/\mathcal{F}(t)$.
\textcolor{black}{The limit of this bound} quantifies the maximum achievable precision $\sigma_t$ for an unbiased estimator measuring $t$. $\mathcal{F}(t)$ can be modelled from analytic probability functions $P_i(t)$ for each outcome ($i$) using
\begin{equation}
\mathcal{F}(t) = \sum_{i=1}^n \frac{1}{P_i(t)}\left(\frac{\partial P_i(t)}{\partial t}\right)^2,
\label{fishereq}
\end{equation}
which highlights \textcolor{black}{the dependence of the maximum precision bound}
on the shape of the interference features given by $P_i(t)$. 
For degenerate photons, $\mathcal{F}(t)$ can be 
increased by broadening the photon spectrum to yield steeper HOM interference features~\cite{Ashley}.
However, increasing the bandwidth of photons leads to a tradeoff between measurement precision and dynamic range, 
\textcolor{black}{while} highly broadband photons can also \textcolor{black}{exacerbate complications arising from} dispersion in the apparatus optics. 
\textcolor{black}{While sensitivity to dispersion is addressed by some terahertz 
pulsed imaging techniques which are widely use in bio-imaging, these can suffer from other issues such as comparatively poor lateral and axial resolution \cite{THZimgcancer,THZimgchina,THZimgkorea}.} 
\textcolor{black}{More broadly, for many high-precision classical imaging techniques, such as super-resolution microscopy, the intensity of illumination required to achieve sub-micron-level precision is between $8$ and $12$ orders of magnitude greater than the microscope proposed in this manuscript \cite{SRM1,SRM2}. Consequently, the HOM microscope can reach a higher Fisher information per photon, which can be an important consideration for photo-sensitive samples and can otherwise be a limiting factor for measurement precision.}

An alternative to increase HOM inteferometer sensitivity is to use the two-colour entangled state~\cite{Raritydip, biphoton}
\begin{equation}
|\Psi\rangle = \frac{1}{\sqrt{2}}\left(|\nu_S\rangle_{_{\mathcal{A}}}|\nu_I\rangle_{_{\mathcal{B}}} +e^{i\phi(t)} |\nu_I\rangle_{_{\mathcal{A}}}|\nu_S\rangle_{_{\mathcal{B}}}\right),
\label{state}
\end{equation}
\textcolor{black}{where $\nu_S$ and $\nu_I$ label frequencies of the signal (horizontally polarised) and idler (vertically polarised) photons}, while $\mathcal{A}$ and $\mathcal{B}$ denote the two separate optical paths that will overlap at a beamsplitter. The resulting two-photon interference of $|\Psi\rangle$ exhibits a beat note with a frequency $\Delta\nu = \nu_S - \nu_I $ that is dependent on the spectral detuning of signal and idler~\cite{Raritydip}. 
This approach has been shown to directly modify $\mathcal{F}(t)$ and has been applied to detect temperature drifts in optical fibre~\cite{biphoton}. 
To apply two-colour HOM interferometry to \textcolor{black}{depth} imaging we construct a tunable wavelength-entangled photon pair source and combine this with both a raster-scanning microscope inside of a HOM interferometer and a quasi-photon number resolving detection scheme~\cite{HOMallcoins}. 

\begin{figure}
\centering
\includegraphics[scale = 0.3]{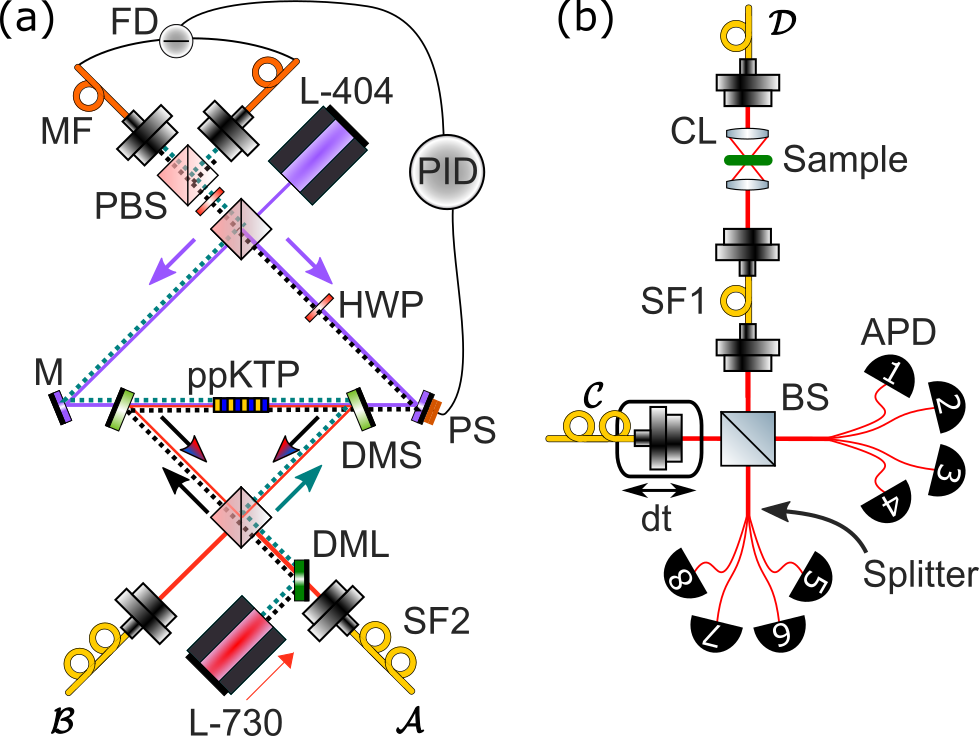}
\caption{\textbf{Schematic of the wavelength-entangled photon pair source and HOM microscope.} (a) Wavelength-entangled photon pair source, comprising: a 404~nm CW pump laser (L-404), polarising beamsplitter (PBS), mirror (M), half waveplate (HWP), periodically poled KTP crystal (ppKTP), piezo-actuated mirror (PS), shortpass dichroic mirrors (DMS), longpass dichroic mirrors (DML), feedback controller (PID), fast balanced photodiode detectors (FD), \textcolor{black}{two-metre length of} 
single mode fibre (\textcolor{black}{SF2}), multimode fibre (MF) and a CW laser at 730~nm to phase lock the source (L-730). $\mathcal{A}$ and $\mathcal{B}$ denote the two output modes from Eq.~\eqref{state}. (b) The HOM microscope, comprising lenses with focal length $f = 11~mm$ in a confocal configuration (CL), 
\textcolor{black}{one-metre length of} 
\textcolor{black}{single mode fibre (SF1)}, 
beamsplitter (BS), $1\times4$ fused fibre coupler (Splitter), time delay for one input beam controlled by a translation stage (dt) and avalanche photodiode detectors (APD). $\mathcal{C}$ and $\mathcal{D}$ denote the input port of the interferometer. Any coincidences between one detector labelled amongst 1 to 4 with one detector labelled amongst 5 to 8 contribute to $N_{11}$, while any coincidences between pairs of detectors amongst 1 to 4 correspond to $N_{02}$ and coincidences between pairs of detectors amongst 5 to 8 give $N_{20}$. This quasi-photon number resolving detection scheme adds an intrinsic loss of $25\%$ for the bunching terms.}
\label{Exps}
\end{figure}

The target entangled state in Eq.~\eqref{state} is generated using a dual-Sagnac arrangement source~\textcolor{black}{\footnote{
\textcolor{black}{This architecture }
\textcolor{black}{was chosen in part to avoid potential delay to a PhD project that that was time-constrained during }
\textcolor{black}{the COVID pandemic.}
}} that needs only one nonlinear down-conversion crystal, as shown in Fig.~\ref{Exps}(a). 
A diagonally polarised CW $404~nm$ laser \textcolor{black}{(linewidth $\Delta\lambda < 5~MHz$; \textit{TOPTICA Photonics - TopMode 405})} is split into two paths by a polarising beam-splitter (PBS) pumping one $3~cm$ long, type-II periodically poled potassium titanyl phosphate (ppKTP) crystal from two directions. 
Down-converted photon pairs centred at $808~nm$ are emitted from the ppKTP crystal in each direction, separated from the pump using shortpass dichroic mirrors (DMS) and subsequently interfere on a second PBS. The spatial mode is filtered with single-mode fibres at the output paths $\mathcal{A}$ and $\mathcal{B}$.

\textcolor{black}{In contrast to polarisation entangled photon pair source designs based on a single Sagnac interferometer \cite{Sagnac,PhaseSagnac} or co-propagating photon source designs based on two crystals in a crossed configuration \cite{biphoton},} the generated photons in this system do not share a common path. \textcolor{black}{As a result,} the source is susceptible to phase instability of the parameter $\phi(t)$ 
\textcolor{black}{of}
the intended output entangled state Eq.~\eqref{state} due to mechanical vibrations and temperature variations.
\textcolor{black}{To counter the sensitivity to phase in generating the required quantum state, we} 
actively stabilise the setup using a diagonally polarised laser at the cut-off wavelength of the DMS (where these are only partially reflective) that is back-propagated through the source optics and monitored with fast balanced 
\textcolor{black}{photodiode}
detectors. The subtracted electric signal from the photodiodes is sent to a PID controller that actuates a piezo stack glued to the mirror of the clockwise pumping path.

The detuning of the central frequencies of the down-converted photons is readily controlled through phase-matching conditions, which in turn can be tuned by changing the crystal temperature whilst leaving the pump laser frequency unchanged. The temperature of the crystal can be set from $30\degree C$ to $180\degree C$ achieving a detuning from $0$ to $30.1~THz$ ($65.6~nm$) 
\textcolor{black}{(see Supplemental Materials). }
The photon pair source generates $\sim200 ~k$ measured coincidences per second with an input pump power of $\sim20~mW$ and we observe a symmetric heralding efficiency up to $25\%$ at the positions marked $\mathcal{A}$ and $\mathcal{B}$ in Fig.~\ref{Exps}(a).

The schematic of the confocal microscope is illustrated in Fig.~\ref{Exps}(b). A semi-transparent sample (with a thickness $d$ and refractive index $n$) is mounted between two lenses in a confocal configuration in path $\mathcal{D}$, before the broadband 50:50 beamsplitter (BS) where the HOM interference occurs. The presence of the sample introduces an additional time delay between paths $\mathcal{C}$ and $\mathcal{D}$ related to the optical path length by $t = {nd}/{c}$. This changes the degree of indistinguishability between the two photons and consequently the relative number of anti-bunching events ($N_{11}$) detected due to the HOM interference effect~\cite{HOMmicroscopy}. Depth features of the sample can also cause path-deviation of the probe photons, thereby altering the spatial overlap of the signal and idler modes at the BS. 
This reduces the visibility of the HOM interference leading to an error in the observed bunching rate and consequently in the sample thickness estimation.
In our experiment we implement single-mode, raster-scanned imaging by translating the sample whilst keeping the probe beam fixed. A single mode fibre after the sample implements spatial mode filtering and ensures a constant overlap of the two spatial modes at the BS.
Any change in detected coincidence rates due to the path-deviation or fluctuations in source brightness is normalised through
the bunching events ($N_{20}$ and $N_{02}$) according to
\begin{equation}
    P_{11} = \frac{\widetilde{N}_{11}}{\widetilde{N}_{11} + \widetilde{N}_{02} + \widetilde{N}_{20}},
    \label{estimate}
\end{equation}
where $\widetilde{N}_{11}$, $\widetilde{N}_{02}$, and $\widetilde{N}_{20}$ are calibrated via the Klyshko efficiency \cite{Klyshko} (see Supplemental Materials).
These values are all dependent on the HOM interference effect and are individually monitored with quasi-photon number resolving detection scheme using multiplexed single photon detectors~\cite{Achilles2004,Xiang2011} as shown in Fig.~\ref{Exps}(b). 
\textcolor{black}{As the HOM effect depends on the detection of coincident photons, all results in this paper estimating the thickness of the sample consider only photons which are post-selected on successful detection.} 
The detectors are silicon single photon avalanche photodiode modules (APD; \textit{Excelitas})  and the coincidences between them are recorded with a Logic16 time-tagger (\textit{UQDevices}).

\begin{figure}[!h]
\centering
\includegraphics[scale = 0.29]{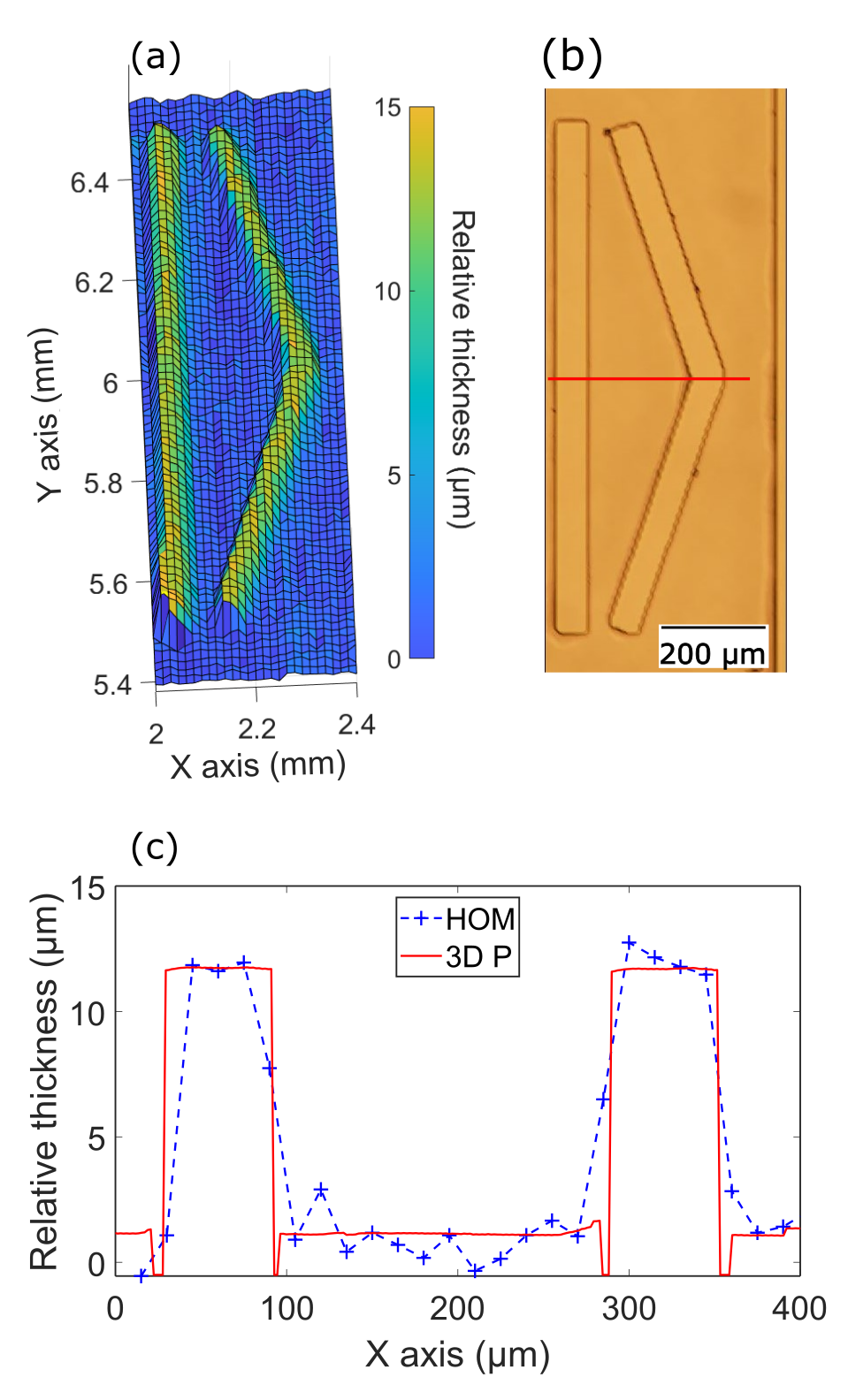}
\caption{\textbf{\textcolor{black}{Depth} sample imaging.} (a) 3D HOM imaging by monitoring the number of coincidences for each pixel. Raster scanning of the sample through the single-mode illumination provides the spatial resolution in the 2D transverse plane. The acquisition time is $0.5~s$ for each pixel ($N\sim 4000$). The frequency beat note is $7.4~THz$, where the detuning is chosen such that the characteristic depth features of this sample correspond to measurements across the full range of half of a fringe period from the HOM interference. This attains the maximum Fisher information while avoiding fringe ambiguity. (b) 2D conventional optical microscope image of the sample \textcolor{black}{with a depth profile of $(10.62\pm0.02)~\mu m$ 
\textcolor{black}{(see Supplemental Materials). }
\textcolor{black}{(c) 1D scan of depth profile following the path indicated by the red line on (b). Results are shown for HOM microscopy (blue) and for a 3D profilometer (red).}} } 
\label{3D}
\end{figure}

\begin{figure}[h]
    \centering
    \includegraphics[scale = 0.27]{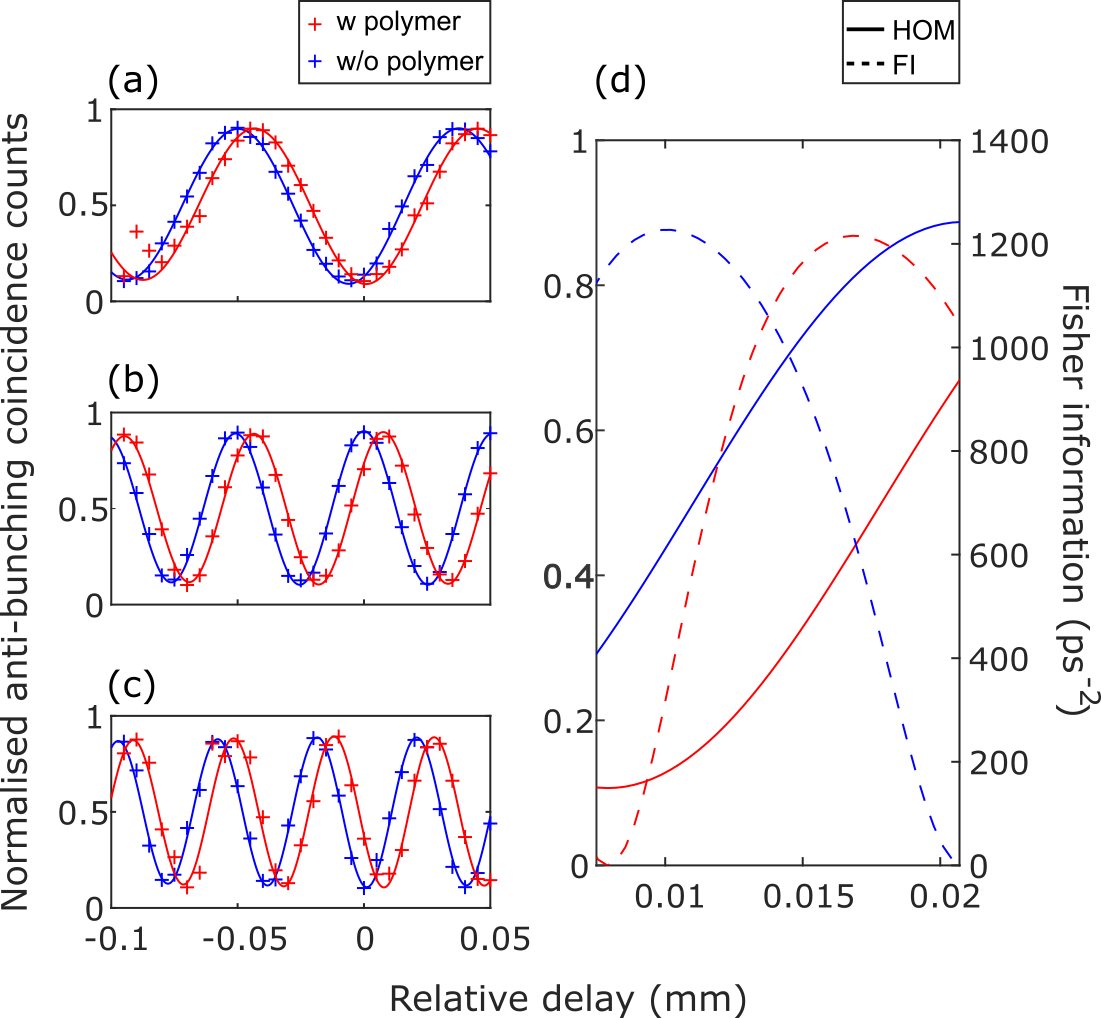}
    \caption{
    \textcolor{black}{\textbf{Timing offset between interference fringes when comparing measurements at two pixels with different sample thickness.} Comparison of the central regions of the HOM interference fringe patterns when applying different frequencies of the beat note: (a) $\Delta\nu = 3.4~THz$, (b) $\Delta\nu=5.9~THz$, (c) $\Delta\nu=7.6~THz$. The blue markers and lines give the results for a selected pixel on the polymer deposition region of the sample, and the red for a pixel in a region with only the substrate and no additional deposited material. Markers correspond to the measured data points, and the lines are a numerical fit to a parameterised model of the interference. (d) Magnified view of a section of (c), with the associated Fisher information per photon shown as dashed lines for both measured positions on the sample. The coarse delay of the reference arm must remain within this range of relative delay to avoid fringe ambiguity. The Fisher information accounts for the intrinsic loss for the bunching terms given by the quasi-photon number resolution scheme used in this experiment.}}
    \label{HOMshifts}
\end{figure}

The transverse resolution of the microscope is limited by the beam waist of the scanning probe, whilst the axial resolution, which is the focus of this work, is given by the frequency of the beat note arising in the HOM interference. 
To maximise the Fisher information for the depth estimate whilst keeping all of the measurements within the same interference fringe half-period (avoiding fringe ambiguity), the signal and the idler photons were detuned by $7.4~THz$ ($16~nm$) giving a \textcolor{black}{spatial} beat note with a half-period $\Lambda/2 = 20.3~\mu m$. For comparison to a classical linear optics approach, performing classical interferometry with an equivalent \textcolor{black}{frequency of oscillation }
would require an electromagnetic radiation source with a central wavelength of $\sim40~\mu m$. Operating at such a wavelength would constrain the transverse spatial resolution due to the diffraction limit. 
\textcolor{black}{On the other hand, classical interference measurements operating at a similar wavelength to the down-converted photons ($\sim 800~nm$) will suffer from phase instability from the environmental background due to their high sensitivity.} 

\begin{figure}[h]
    \centering
    \includegraphics[scale = 0.25]{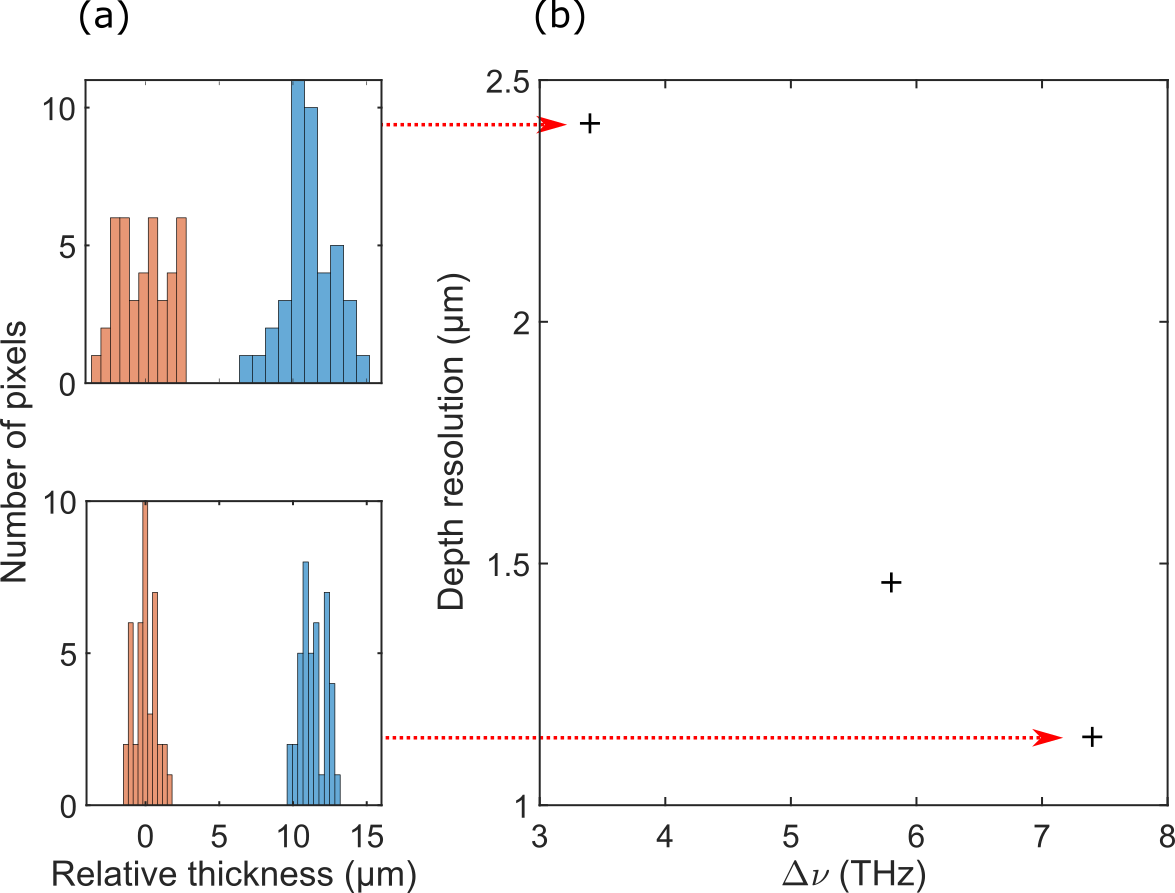}
    \caption{\textbf{Axial precision for varying beat note frequency.} (a) Histograms of the axial depth estimate for the two step measurement for beat note frequencies of 
    \textcolor{black}{$3.4~THz$ }
    \textcolor{black}{(top)} and $7.4~THz$ \textcolor{black}{(bottom)}. (b) Precision of depth estimation calculated from the raw histogram data using Eq.~\eqref{sigma} 
    \textcolor{black}{after correction for the refractive index of the sample}.}
    \label{Precision}
\end{figure}

A test sample for imaging was fabricated by depositing micron-scale SU-8 polymer features with a thickness \textcolor{black}{$(10.62\pm 0.02)~\mu m$} onto a silica substrate 
\textcolor{black}{(see Supplemental Materials)} 
in the shape of a \enquote{KET} symbol (shown in Fig.~\ref{3D}(b)). \textcolor{black}{The thickness variation of the sample on a 2D scan} is shown in Fig.~\ref{3D}(a).

\textcolor{black}{For initial calibration of the HOM microscopy system, the HOM interference features were measured for two pixels positions on the sample (with and without deposited polymer) by translating the coarse delay in the reference arm (Fig.~\ref{HOMshifts}). Both measurements showed interference fringes with a visibility of $\sim 80\%$.}
\textcolor{black}{The sample induces a relative time delay between the reference and the probe photons, leading to an offset of $\sim 8~\mu m$  between the two interferograms. Assuming a sample refractive index of $n=1.58$ for the deposited structure, this corresponds to a measured thickness change of $\sim 14~\mu m$.}
\textcolor{black}{For acquiring a full sample image this approach would be prohibitively slow, due to the requirement of repeating the scan of a HOM interference fringe for every pixel. Instead, the coarse delay is set at a fixed position within the range of $0.01$ to $0.02~mm$ to avoid fringe ambiguity and the normalised anti-bunching coincidence counts are monitored for each pixel.} 
\textcolor{black}{The beat note is chosen to optimise the width of the Fisher information peak such that fringe ambiguity is avoided. Subsequently, the coarse delay is set at the crossing point for the Fisher information curves for the maximum and the minimum required depth measurement values. This provides a compromise by optimising Fisher information and precision at both extremes of the measurement range.}
\textcolor{black}{Finally, the relative time delay induced by the variation of the thickness within the sample is estimated through the variation in the probability outcomes $N_{11}$.}

\textcolor{black}{The phase-locking was monitored with an}
\textcolor{black}{electronic data logger}
\textcolor{black}{(\textit{Liquid Instruments - Moku:GO}) to ensure stability of the photon pair source during the data acquisition (see Supplemental Materials).}
\textcolor{black}{If failure in the phase-locking is detected, the system automatically suspends data acquisition and reacquires locking with a new set point for the PID controller. As this results in an unknown offset in the absolute value of the measured sample thickness, it is also necessary in this case to repeat the measurement of the previously acquired pixels.} 
$27 \times 78$ pixel positions were measured by monitoring the outcomes of the HOM interferometer, with data acquisition time $\sim20~min$ ($\sim40~min$ total measurement time due to the limited speed of the translation stages) for an area of $0.4~mm^2$. The maximum transverse resolution of the HOM microscope was determined to be \textcolor{black}{$\sim 10\times10~\mu m^2$} with a resolution test target, but due to the size of this sample, a reduced transverse resolution ($\sim15\times15~\mu m^2$) was used to decrease the total acquisition time. 
\textcolor{black}{Fig.~\ref{3D}(c) shows a comparison of a 1D raster scan across a single row of pixels using the HOM microscope with the results from a commercial 3D profilometer} (\textcolor{black}{\textit{Filmetrics - Profilm3D$^\circledR$}}).
\textcolor{black}{\textcolor{black}{As the raster-scan pattern consisted of a vertical sweep in the Y axis followed by stepwise increment on the X axis, the 1D scan results correspond to data points monitored over a period of $\sim40~min$. The results from the HOM microscope agree with the measurements from the 3D profiler (within the given error of $1.1~\mu m$ of the HOM microscope system) across the whole time period of the measurement. T}he sharpness of the edge features observed in the sample profile were limited for the HOM microscope measurement by the reduced transverse resolution.}

Assuming the known refractive index value for this polymer to be $n=1.58$ \textcolor{black}{for both the signal and idler} 
\textcolor{black}{wavelengths, }
the measured relative thickness between the two steps in this sample is determined to be \textcolor{black}{$11\pm 1~\mu m$}. 

Fig.~\ref{Precision} shows the results of repeated measurements of the sample for different beat note frequencies, which was used to characterise the axial precision. For each parameter combination, 
\textcolor{black}{depth estimates were made by acquiring and combining data from groups of the same 40 pixels (a subset of a pixel column in the image). This was performed on two distinct regions of the sample - with one corresponding to pixels lying on the deposited polymer, which we denote as step 1 ($S_1$), and the other on a region where no polymer was deposited, which is denoted as step 2 ($S_2$).} 
As the depth estimates for the two regions are statistically independent, the overall precision combining the two data sets is given by 
\begin{equation}
    \sigma_t = \sqrt{var(S_1)+var(S_2)}.
    \label{sigma}
\end{equation}

Fig~\ref{Precision}(a) shows examples of the underlying histogram results of the measured depth for two of the data points in Fig~\ref{Precision}(b). The effect of increasing $\Delta\nu$ is evident from the reduced variance of the data set of $S_1$ and $S_2$ arising from the increase in the Fisher information. Consequently the precision of the measurements is improved and smaller depth features can be more easily distinguished. The best precision result of 
\textcolor{black}{$0.8~\mu m$ for the relative sample thickness of each of the two steps individually, leads to a combined precision of $1.14~\mu m$ for the estimate of the relative depth offset between the two steps by using Eq.~\eqref{sigma}.}

\begin{figure}[h]
    \centering
    \includegraphics[scale = 0.25]{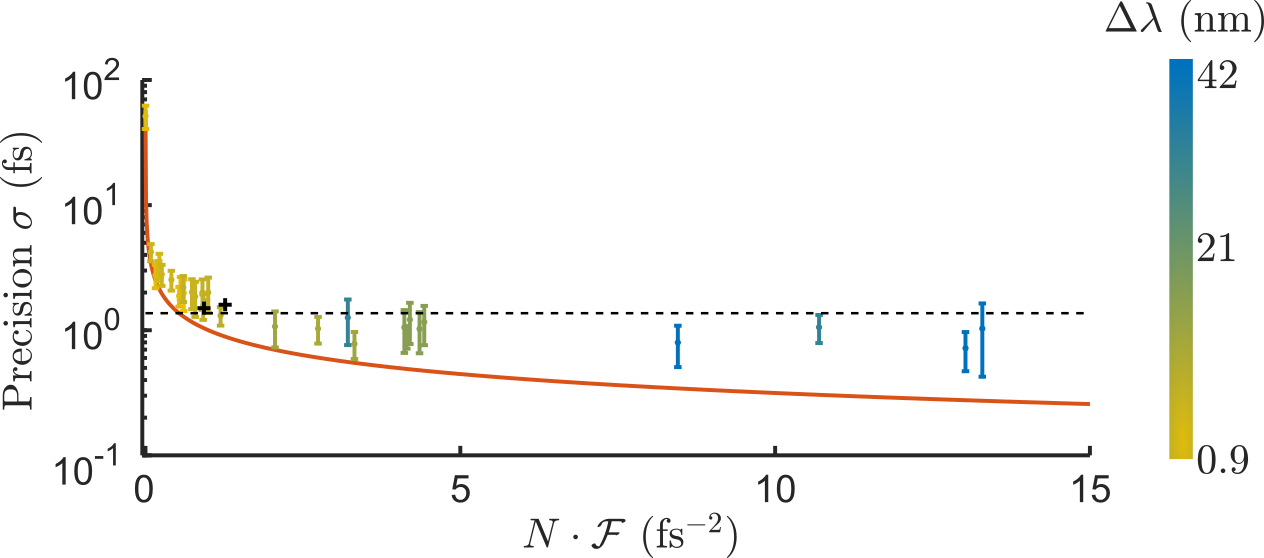}
    \caption{
    \textcolor{black}{\textbf{Dependence of measured axial precision on total of Fisher information for single pixel estimation.} Total Fisher information is given by the product of the number of detected photon pairs $N$ and the theoretical Fisher information per photon $\mathcal{F}$ (Eq.~\eqref{fishereq}). Fisher information is controlled by varying the wavelength separation $\Delta\lambda$ of the downconverted photons 
    (colour bar on right side)}. 
    The acquisition time is fixed to $0.5~s$ and the number of \textcolor{black}{detected} photon pairs to $N\sim10^4$ for every data 
    \textcolor{black}{point}. Each point on the graph corresponds to 500 independent time delay measurements split into blocks of 50, where the mean of the standard deviation of these blocks corresponds to the precision. The standard deviation from these 10 blocks gives the error bars. The orange line corresponds to the theoretical bound of the precision for $N$ photons \textcolor{black}{saturating} the Cram\'{e}r Rao bound (CRB), and the \textcolor{black}{two black crosses correspond} to the 
    \textcolor{black}{two step measurement from} 
    Fig.~\ref{3D}(a)\textcolor{black}{. The horizontal dashed line is the}
    \textcolor{black}{threshold for achieving }
    \textcolor{black}{sub-$\mu m$ axial depth precision ($\sigma<1.4~fs$)}\textcolor{black}{, given by }
    \textcolor{black}{Eq.~\eqref{sigma} for a sample with a refractive index $n=1.58$}.}
    \label{NFI}
\end{figure}
%
%
In order 
to decouple thickness variation in the sample from the performance of the microscope, repeated measurements were performed on a single pixel. 
\textcolor{black}{The normalised anti-bunching coincidence counts }
\textcolor{black}{were }
\textcolor{black}{measured $500$ times for different }
\textcolor{black}{frequencies } 
\textcolor{black}{of the beat note (from $0.09~THz$ to $18~THz$). For each of the }
\textcolor{black}{data points, }
\textcolor{black}{the $500$ measurements are split into $10$ blocks of $50$ measurements. The standard deviation of these $10$ blocks is calculated, giving a new data set.}
\textcolor{black}{The mean of each grouped data set gives the precision value of the corresponding data point shown in Fig.~\ref{NFI} and the error bars are given by the standard deviation of the grouped data sets. }
In our lab the measurements were stable over a time period of \textcolor{black}{$\sim40~min$} and drift in the phase occurred over longer periods of time, which resulted from mode-hopping instability of the laser used to phase-lock the source. Changes in the output wavelength of the reference laser lead to fluctuations in the phase-locking conditions in the interferometer and drifts in the depth estimation. We also observe that as the Fisher information increases, the measured precision deviates from the CRB. We believe that this is due to an additional phase-locking instability observed at high crystal temperatures ($\geq120~\degree C$) \textcolor{black}{causing perturbations in the refractive index.} Two-colour photon pair sources designed to be more robust to phase instability may provide an improvement in the saturation of the CRB for higher frequency beat notes~\cite{biphoton}.

Using Eq.~\eqref{sigma}, it can be shown that sub-$\mu m $ precision can be achieved for a measurement with total Fisher information 
($N\cdot\mathcal{F}$) \textcolor{black}{above $ 0.5~fs^{-2}$. As shown on Fig.~\ref{NFI}, this value is obtained with $N=10^4$ \textcolor{black}{detected} photon pairs and a detuning of $12.3~nm$. The black crosses show the theoretical Fisher information and the precision achieved for each of the step measurements from Fig.~\ref{3D}(a). Therefore, sub-$\mu m$ precision for depth imaging would be expected for a source with the same detuning and twice the brightness.}

The total loss in the source and the microscope was relatively high ($\sim98\%$ for each photon). This included $6~dB$ of loss for FC to FC fibre connectors, which could be removed. This limited the total amount of Fisher information by reducing the number of detected probe photons for a given acquisition time. Reducing the loss and optimising the brightness of the source could improve the axial resolution by an order of magnitude. To increase the precision further on more complex samples with greater thickness variations, whilst avoiding fringe ambiguity, one could perform an initial coarse mapping of the depth profile using raster-scanning with a low frequency beat note, followed by a scan at increased resolution with a larger wavelength detuning. High resolution scanning could also be reserved for targeted areas of interest to reduce the required measurement time.

We have shown that HOM interference can be implemented with two-colour entangled photon pairs \textcolor{black}{to achieve sub-$\mu m$ precision for depth estimation (approaching the performance of super resolution microscopy) and with a probe light intensity of $10^{-8}~W\cdot ~cm^{-2}$ incident on the sample. The dynamic range offered by} measurement techniques of this type \textcolor{black}{combined with sub-$\mu m$ axial precision} could be particularly relevant for characterising in situ, sub-cellular structures for semi-transparent samples such as biological cells.

\section*{Acknowledgements}
The authors are grateful to M.~Piekarek, G.~Ferranti, A.~Belsley and J.~Rarity for helpful discussions. This work was supported by Engineering and Physical Sciences Research Council funded projects Nano-HOM (EP/R024170/1) QuantIC - The UK Quantum Technology Hub in Quantum Imaging (EP/T00097X/1). CT was funded by the EPSRC CDT in quantum engineering (EP/L015730/1). JM was supported by ERC starting grant PEQEM (803665). JMR would like to acknowledge funding from the EPSRC New Horizons project EP/V048856/1.

\newpage
.
\newpage
\section{Supplemental Materials}
\subsection{Hong-Ou-Mandel for two-colour state}
With a type-II crystal and unfiltered, degenerate photon pair source, the probability of the anti-bunching state at the output from the HOM interferometer is modelled by~\cite{Multiphoton}:
\begin{equation}
    P_{11}(t)=\left\{
    \begin{array}{ll}
    1/2\left[1-\alpha \left(1-\left|\frac{t}{\tau}\right|\right)\right],&\text{for}~\left|t/\tau\right| \leq 1  \\
    1/2,&\text{for}~\left|t/\tau\right| > 1\end{array}\right.
    \label{eqHOM}
\end{equation}
where $\alpha$ is the visibility of the HOM interference \textcolor{black}{and is strongly dependent on the quantum state visibility}, $t$ is the relative time delay between the two interfering photons and $\tau$ is the temporal width of the photons. Due to the type of the crystal and the absence of narrowband spectral filtering within the experimental setup, the downconverted photons ideally match the temporal shape of a top-hat function, with   $\tau = DL$, where $L$ is the length of the crystal and $D$ is the difference between the inverse group velocities of the two downconverted photons~\cite{Vshape1,Vshape2}.

To model the HOM interference of the two-colour entangled photons given by:
\begin{equation}
    |\Psi\rangle = \frac{1}{\sqrt{2}}\left(|\nu_S\rangle_{_{\mathcal{A}}}|\nu_I\rangle_{_{\mathcal{B}}} +e^{i\phi(t)} |\nu_I\rangle_{_{\mathcal{A}}}|\nu_S\rangle_{_{\mathcal{B}}}\right),
    \label{supstate}
\end{equation}
we multiply the term $\alpha\left(1-|t/\tau|\right)$ from Eq~.\eqref{eqHOM} by an additional term $cos(2\pi\Delta\nu\tau+\phi(t))$ giving
\begin{equation}
    P_{11}(t) = \frac{1}{2}\left[1-\alpha \left(1-\left|\frac{t}{ \tau}\right|\right)cos(2\pi\Delta\nu\tau+\phi(t))\right].
    \label{eqbiphoton}
\end{equation}
$\Delta\nu = \nu_S - \nu_I$ is the frequency detuning between the signal and the idler and $\phi(t)$ is the relative phase between the two sub-states from Eq.~\eqref{supstate}. Due to conservation of energy, the probability of the bunching terms $N_{02}$ and $N_{20}$ is given in the ideal case by:
\begin{equation}
    P_{20} = P_{02} = \frac{1-P_{11}}{2}.
\end{equation}

\subsection{Calibration with the use of Klyshko efficiency}
The calibration of $N_{11}$, $N_{20}$ and $N_{02}$ is implemented through the Klyshko efficiency \cite{Klyshko}. 
We utilise the measured anti-bunching rate at a sufficiently large time delay to avoid HOM interference and define $\eta_i$ to be the Klyshko efficiency of the $i^{th}$ channel ($i=1,2,3,4$),
\begin{equation}
    \eta_i = \frac{1}{4}\sum_{j=5}^8 \frac{C_{ij}}{2/3 \cdot Sj}.
\end{equation}
$C_{ij}$ are the raw coincidence counts between the detectors $i$ and $j$. $S_{j}$ corresponds to the single counts detected by detector $j$ ($j=5,6,7,8$). With $P_{11} = \frac{1}{2} = 2\times P_{20} = 2\times P_{02}$, only $\frac{2}{3}$ of the single counts from the detectors $j$ will correspond to the state $N_{11}$ while $\frac{1}{3}$ of these single counts belong to the anti-bunching state $N_{20}$. Likewise, similar expressions can be derived for the other detector channels, where the roles of i and j are reversed.
Finally, each coincidence count total $C_{ij}$ can be normalised using the corresponding heralding efficiency coefficients to account for unbalanced losses.

\subsection{Experimental characterisations}
\begin{figure}[!h]
    \centering
    \includegraphics[scale = 0.3]{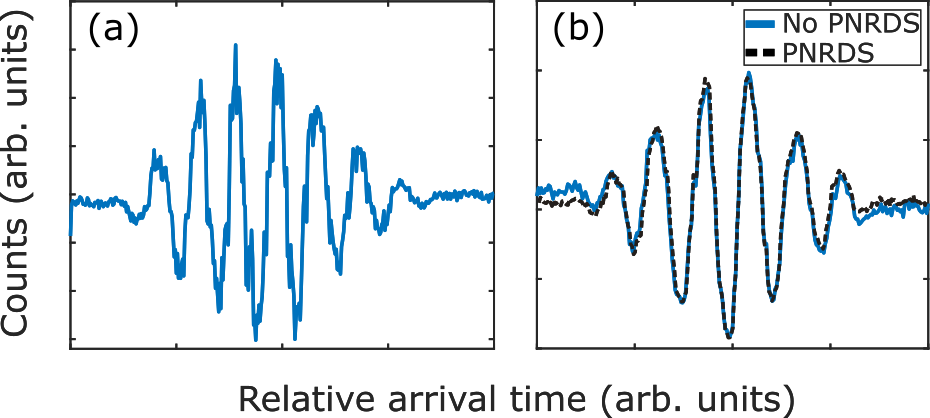}
    \caption{\textbf{Example of HOM interference stabilisation.} (i) No phase-locking or normalisation, (ii) phase-locking and no normalisation (blue), phase-locking and normalisation with the use of PNRDS (dashed black). }
    \label{HOMlocks}
\end{figure}

The wavelength-entangled photon pair source has to be actively phase-locked to remove instability in the HOM interference, which otherwise results in noise in the measurement as shown in Fig.~\ref{HOMlocks}(a). 
In order to distinguish between changes in the brightness of the source and changes in the loss or bunching rate induced by the sample, the use of quasi-photon number resolving detection scheme (PNRDS) allows the bunching and anti-bunching terms to be monitored independently, normalising the measurement. The effect of imbalanced alignment is apparent in Fig.~\ref{HOMlocks}(b) from asymmetry in the baseline count rate on the extreme right-hand and left-hand sides of the image (blue line), and is corrected by the use of the PNRDS (dashed black line).

\begin{figure}[!h]
    \centering
    \includegraphics[scale = 0.35]{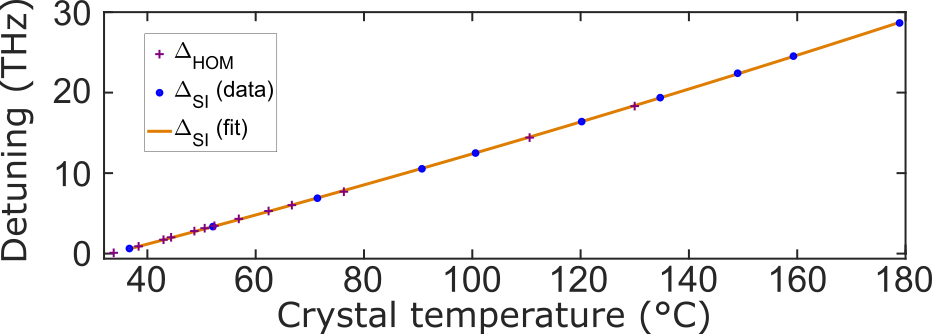}
    \caption{\textbf{Detuning of the photon pair.} Blue data points are the experimentally measured values for detuning of the signal and idler peaks at different crystal temperatures. 
    $\Delta\nu_{_{SI}}$ (yellow line) is a quadratic fit to these data points. 
    $\Delta\nu_{_{HOM}}$ (purple crosses) shows expected values for the signal and idler detuning. These were calculated using Eq.\eqref{eqbiphoton} from the frequency of the beat note of a measured HOM dip at each corresponding crystal temperature.}
    \label{DetuningVStemp}
\end{figure}
Fig.~\ref{DetuningVStemp} shows characterisation data for the frequency detuning between the signal and the idler of the photon pair source. The central wavelengths of the signal and idler peaks were measured with a sensitive, high resolution spectrometer (\textit{Andor}, \textit{SR-750-B2-R-SIL}) for a range of crystal temperatures. 

\textcolor{black}{The pump laser wavelength was measured to be $\lambda_p = 404.093~nm$ with a linewidth of $\Delta\lambda_{p} = 5~MHz$ (Topmode 405, Toptica).}
As shown in Fig.~\ref{DetuningVStemp}, the detuning values predicted using the HOM interference beat note show good agreement with the calibrated detuning curve measured using the spectrometer. 



\subsection{Sample preparation}
The fabricated semi-transparent sample consisted of a 1~mm thick fused silica substrate where a $\sim 10~\mu m$ thick film of SU-8 polymer was deposited by spin-coating. SU-8 is a negative epoxy photoresist with a high optical transparency making it ideal for this work. Using photolithography, an image of a "KET" was transferred to the polymer, resulting in a semi-transparent sample with \textcolor{black}{a variation thickness of $(10.62 \pm 0.02)~\mu m$}.


\end{document}